\documentclass[nolinenumbers]{aastex631}

\submitjournal{PSJ}

\shorttitle{Tidal Evolution with High Obliquity}
\shortauthors{\'Cuk et al.}

\graphicspath{{./}{figures/}}

\begin{document}

\title{Tidal Evolution of the Earth-Moon System with a High Initial Obliquity}

\correspondingauthor{Matija \'Cuk}
\email{mcuk@seti.org}

\author[0000-0003-1226-7960]{Matija \'Cuk}
\affiliation{SETI Institute \\
189 N Bernardo Ave\\
Mountain View, CA 94043, USA}

\author{Simon J. Lock}
\affiliation{Division of Geological and Planetary Sciences \\
California Institute of Technology\\
1200 East California Boulevard\\
Pasadena, CA 91125, USA}

\author{Sarah T. Stewart}
\affiliation{Department of Earth and Planetary Sciences\\
University of California Davis\\
One Shields Avenue\\
Davis, CA 95616, USA}

\author{Douglas P. Hamilton}
\affiliation{Department of Astronomy\\
University of Maryland\\ 
Physical Sciences Complex\\ 
College Park MD 20742, USA}



\begin{abstract}

A giant impact origin for the Moon is generally accepted, but many aspects of lunar formation remain poorly understood and debated. \citet{cuk16} proposed that an impact that left the Earth-Moon system with high obliquity and angular momentum could explain the Moon's orbital inclination and isotopic similarity to Earth. In this scenario, instability during the Laplace Plane transition, when the Moon's orbit transitions from the gravitational influence of Earth's figure to that of the Sun, would both lower the system's angular momentum to its present-day value and generate the Moon's orbital inclination. Recently, \citet{tia20} discovered new dynamical constraints on the Laplace Plane transition and concluded that the Earth-Moon system could not have evolved from an initial state with high obliquity. Here we demonstrate that the Earth-Moon system with an initially high obliquity can evolve into the present state, and we identify a spin-orbit secular resonance as a key dynamical mechanism in the later stages of the Laplace Plane transition. Some of the simulations by \citet{tia20} did not encounter this late secular resonance, as their model suppressed obliquity tides and the resulting inclination damping. Our results demonstrate that a giant impact that left Earth with high angular momentum and high obliquity ($\theta > 61^{\circ}$) is a promising scenario for explaining many properties of the Earth-Moon system, including its angular momentum and obliquity, the geochemistry of Earth and the Moon, and the lunar inclination.
\end{abstract}


\section{Introduction} \label{sec:intro}

The Moon is widely thought to have formed in the aftermath of a giant impact between the proto-Earth and another planetary body close to the end of terrestrial planet formation \citep{har75, cam76, can01, asp14, bar16, loc20a}. The original Giant Impact hypothesis has been built on the  classical reconstructions of past tidal evolution of the Earth-Moon system \citep{gol66, tou94}, which indicated that when the Moon first formed, at a distance of a few Earth radii, Earth had a 5-hour spin period and an obliquity of about 10$^{\circ}$. The giant impact that produced this state was inferred to involve a grazing, Mars-sized impactor with a very small relative velocity at infinity \citep{can01}, and this giant impact scenario is commonly known as the "canonical" model. Numerical simulations have found that a canonical impact results in a Moon made predominately from the material from the mantle of the impactor \citep{can01}. However, the isotopic signature of the Moon appears to be very similar to that of Earth \citep{wie01, zha12, you16, Touboul2015, Kruijer2015}, implying a common source for terrestrial and lunar materials, and leading to claims of an "isotopic crisis" for lunar formation \citep{mel09}.  

This isotopic crisis inspired new modifications to the Giant Impact hypothesis, notably the idea that the post-impact Earth-Moon system had much higher angular momentum (AM) than it has now \citep{cuk12}.  Higher AM impacts may be able to produce an Earth and Moon with similar mixtures of impactor and target material and consequently similar isotopic signatures \citep{cuk12, can12, loc18}. Chemical equilibrium under greater gas pressures within circumterrestrial disks generated by high-energy, high-AM impact events may also explain the relative deficits of moderately volatile elements in the Moon compared to Earth \citep{loc18}. To be consistent with the current Earth-Moon system, the high-AM hypothesis requires that some dynamical mechanism subsequently transfer AM away from the Earth-Moon system. Capture into the evection resonance was the first such mechanism to be proposed \citep{cuk12}, but requires high lunar eccentricity which may be hard to maintain \citep{tia17, war20}; near-resonant interactions with evection now appear to be a more robust option \citep{wis15, tia17, Rufu2020}. 

Independent of the issue of lunar composition, classical reconstructions of lunar tidal evolution were incomplete, as they did not include obliquity tides, which may have greatly reduced lunar inclination during the Moon's early history \citep{che16}. The lunar orbital inclination (currently 5$^{\circ}$), was thought to be about 12$^{\circ}$ soon after lunar accretion \citep{gol66, tou94}, but the inclusion of obliquity tides suggests a much greater early orbital tilt for the Moon. \citet{cuk16} proposed that tidal evolution of the Moon from a high-obliquity, high-AM Earth could result in both a large early lunar inclination (about 30$^{\circ}$) and the transfer of AM from the Earth-Moon system to the Earth-Sun system. These dynamical effects are due to instability of the orbits during the Laplace Plane transition (LPT -- the orbital zone where the solar and terrestrial oblateness-driven perturbations are comparable) for satellites of high obliquity planets \citep{tre09}.

Recently, \citet{tia20} have shown that the Laplace Plane instability, like other solar interactions with the Earth-Moon system, should conserve the component of the AM vector of the system that is perpendicular to the ecliptic (referred to by \citet{tia20} as the "vertical AM" but which we will refer to as "ecliptic AM"). The simulations of \citet{cuk16} did not conserve this quantity, casting serious doubt on the results of \citet{cuk16}, including the Earth-Moon system AM loss through the LPT. Beyond the issue of ecliptic AM conservation, \citet{tia20} stated that evolution through the Laplace Plane instability cannot result in the current Earth-Moon system regardless of the tidal parameters assumed for the system. \citet{tia20} found that all high-obliquity Earth-Moon pairs with the correct ecliptic AM led to systems with too much AM and too high an obliquity for Earth at the end of the LPT. 

Here we revisit the work of \citet{cuk16} in light of the findings of \citet{tia20}. We correct and improve the numerical integrator developed by \citet{cuk16}, and use it to explore the dynamics of an Earth-Moon system with a high initial AM and obliquity, and determine whether this evolution can result in the today's observed configuration.

\section{Numerical Integrator}

In this work we used a corrected and improved version of the {\sc r-sistem} integrator, which we will refer to as {\sc r-sistem8}. Orbital motions are integrated using a mixed-variable symplectic approach \citep{wis91}, with the solar perturbations included using the algorithm of \citet{cha02}. Rotational motion of the triaxial Moon was integrated using the Lie-Poisson approach of \citet{tou94lp}, while Earth is assumed to be oblate and in principal axis rotation. Here we will briefly describe the differences between the new version of the integrator and that used by \citet{cuk16}.

\begin{figure}
\epsscale{.6}
\plotone{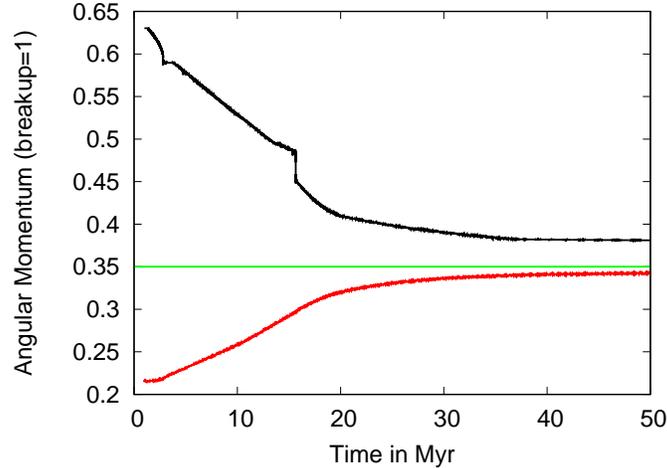}
\caption{Variation in the total (top line in black) and ecliptic AM (lowest line in red) in the simulation shown in Fig. 1 of \citet{cuk16}. The current AM of the Earth-Moon system is plotted by the middle line in green. In the absence of the erroneous obliquity drift, the ecliptic AM should be approximately constant. The fact that the red line is not constant illustrates the error in our earlier approach in \citet{cuk16}. The unit of AM is $L_C=\alpha \sqrt{G M^3 R}$}
\label{am_cons}
\end{figure}

\begin{figure}
\epsscale{0.6}
\plotone{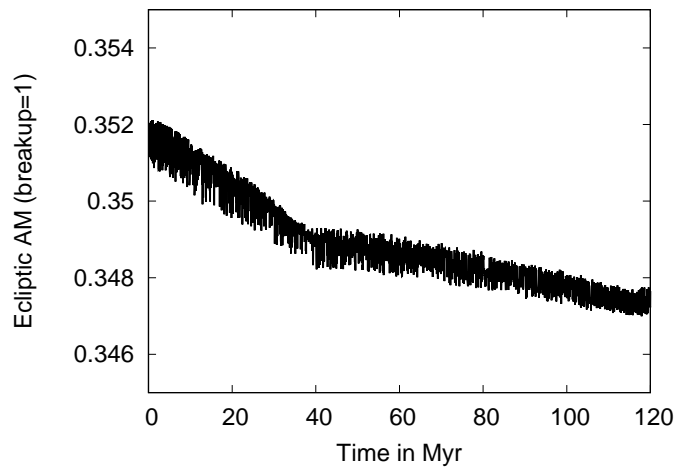}
\caption{Variation of the ecliptic AM in the simulation shown in Fig. \ref{rev8}. Errors are at the $\approx 1\%$ level in contrast to Fig. \ref{am_cons}. The unit of AM is $L_C=\alpha \sqrt{G M^3 R}$.}
\label{new_am}
\end{figure}

Incorrect handling of the axial precession of Earth was the main source of error in the results of \citet{cuk16}. Their integrator simply gave the Earth's axis a single ``kick'' per timestep (calculated using the positions of the Sun and the Moon), in a leapfrog scheme where perturbations were alternating with the Keplerian motions. While this approach did not lead to any secular errors when integrating the current Earth-Moon system or during the lunar Cassini state transition, significant drift in obliquity occurs when the Moon is in one of the secular resonances observed during the LPT and so usual symmetries do not apply. Since our integrator conserved the total spin of Earth, but kept artificially lowering the obliquity, this led to a secular increase in the ecliptic angular momentum. Figure \ref{am_cons} shows that about a third of the total change in AM in \citet{cuk16} Figure 1 was due to this ``obliquity leakage''. We corrected this problem by replacing this simple mapping by a fourth-order Runge-Kutta step \citep{pre92} which approximates the precession of Earth under lunar and solar torques. Unlike our treatment of orbital motions and lunar rotation, this approach is not symplectic, so absolute conservation of ecliptic AM is not expected. However, ecliptic AM is now approximately conserved and its variation in our simulations of the Laplace plane transition (LPT) is approximately 1\% (Fig. \ref{new_am}).   

\citet{cuk16} assumed that Earth's oblateness moment depends on its spin rate $\omega_R$ as $J_2 \propto \omega_R^2$. While this is accurate at low rotation rates, early Earth is significantly deformed in some high-AM scenarios \citep{cuk12, loc20} and so we cannot assume oblateness is a small correction to the planet's figure. Based on the results of \citet{loc20}, we now use the relationship $J_2 \propto L^2 (1-0.64 (L/L_C))$ where $L$ is the Earth's rotational AM, and $L_C=\alpha \sqrt{G M^3 R}$ is the critical AM at which the equator of a (non-deformable) Earth would reach orbital velocity ($M$, $R$ and $\alpha$ are Earth's present-day mass, radius and dimensionless moment of inertia, and $G$ is the gravitational constant). Note that here we are using the Earth's AM rather than its spin rate, as our precession routine only needs to ``know'' about $J_2$ and $L$, and we do not need to track the spin rate explicitly. The fact that we encounter the Laplace Plane instability at the same distances as \citet{tia20} indicates that our approaches yield approximately similar results.

\citet{cuk16} designed their integrator to model the Cassini State transition of the lunar spin axis, which occurs later in lunar tidal evolution than the dynamical events studied in this work. Therefore, the earlier version of {\sc r-sistem} assumed a constant figure for the Moon. This is unlikely to be true during the earliest stages of lunar tidal evolution, when the Moon may have still had a magma ocean, and the lunar figure was likely in (or close to) hydrostatic equilibrium. While \citet{cuk16} periodically manually adjusted the lunar figure in their simulations of the LPT, here we implement automatic adjustment of the Moon to the triaxial figure calculated by \citet{kea14} at the relevant semimajor axis (assuming zero eccentricity and obliquity). As the Lie-Poisson approach we adopted \citep{tou94lp} assumes a rigid-body Moon, we tried to minimize the number of figure adjustments, so we included it into a routine activated about every 1000~yr. Therefore, we adjusted the lunar figure about $10^5$ times in the simulations shown in Figs. 1 and 3, which is only a small fraction of about $2 \times 10^{11}$ steps taken by the integrator. To make sure that our reshaping events do not lead to leakage of AM or energy, we re-ran a part of the simulation shown in Fig. \ref{rev8} with a 10,000~year interval (rather than 1000~yr interval) of adjusting lunar shape and found no noticeable change in outcome. 

We use essentially the same tidal model as \citet{cuk16}. The similarity of our results to those of \citet{tia20} suggests that our fundamentally different approaches produce consistent outcomes. However, in their supplementary discussion, \citet{tia20} make several incorrect statements about our tidal model that we need to clarify. Our tidal model is based on instantaneous motions, both angular and radial, of Earth, the Moon and the Sun relative to each other. We do not use orbital elements in calculating tidal accelerations. This choice was made so we would not have to make any assumptions about lunar rotation, which can become asynchronous during the Cassini State transition \citep{cuk16, cuk19}. Therefore, when modeling lunar librational eccentricity tides, we use the instantaneous rotational motion of the Moon relative to the Earth-Moon line as the main input parameter. When calculating tidal despinning of a non-synchronous satellite, the ``constant Q'' approach means that the tidal bulge is assumed to be at a fixed angular distance $\epsilon$ from the sub-planet point, with synodic rotation rate of the planet determining only the sign of this angle, but not its magnitude $\epsilon = (2Q)^{-1}$ \citep{md99}. Applying the same approach to lunar librations would make the amplitude of librations (proportional to eccentricity in absence of other dynamical excitation) irrelevant for the resultant force on Earth and the opposing torque on lunar figure. In this situation, the magnitude of eccentricity damping  would be independent of eccentricity, which is not physically correct. Because of this, our integrator assumes linear frequency dependence of the tidal response for rotational rates comparable to the mean motion, as described in \citet{cuk16}. The fact that we cannot use a constant-$Q$ approach to calculate the lunar librational motion is not in contradiction with the more traditional approach of \citet{tia20}, who use a constant $Q$ model of tides which assumes synchronous rotation and explicitly involves lunar orbital eccentricity. 

\section{Dynamical Stages of the Laplace Plane Instability} \label{sec:lpi}

\begin{figure}
\plotone{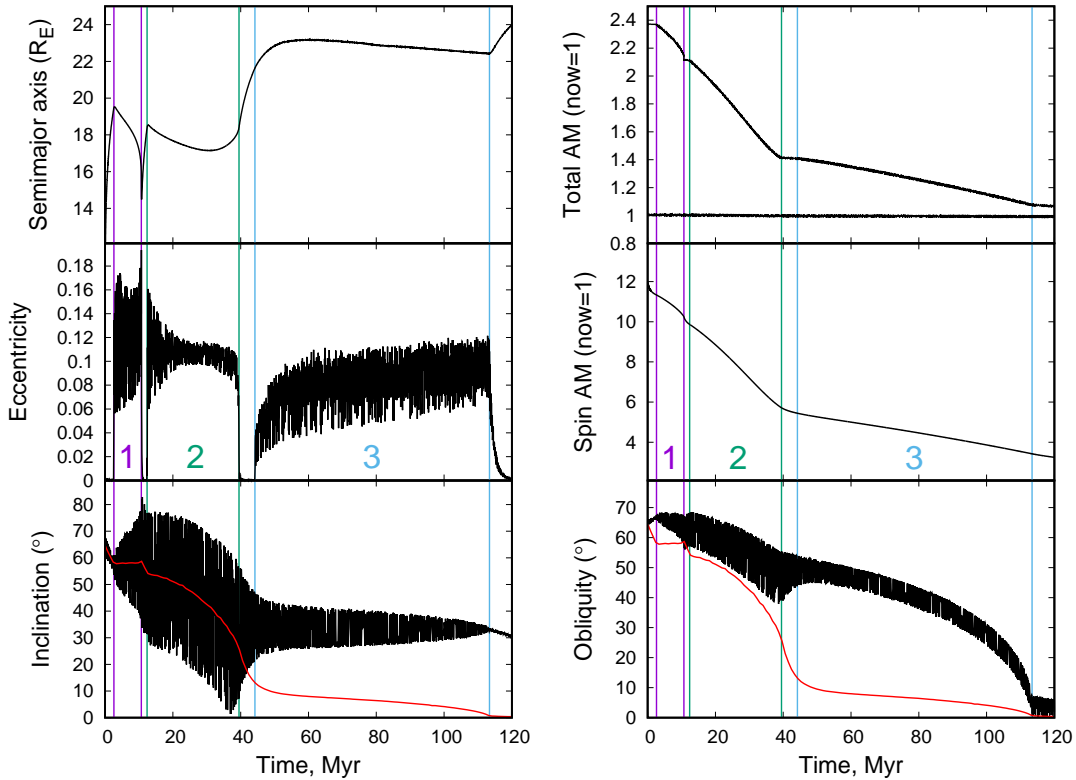}
\caption{Evolution of the Earth-Moon system starting with a 65$^{\circ}$ obliquity and 2.37 times its present-day total AM (chosen to match the present ecliptic AM, $L=0.35 L_C$; see supplementary text). Tidal parameters were set to $Q_M/k_{2M}=Q_E/k_{2E}=100$, and the figures of Earth and the Moon were assumed to be in hydrostatic equilibrium. The left-hand panels plot (top to bottom) the lunar semimajor axis, eccentricity, and inclination with respect to the ecliptic. Also top to bottom, the right hand panels plot the system's AM (solid curve is total AM, nearly-horizontal line ecliptic AM), spin AM of Earth (roughly equivalent to number of rotations per day), and Earth's obliquity relative to the ecliptic. The solid red lines plot the averaged tilt to the ecliptic of the lunar Laplace plane \citep[][Eq. 22]{tre09}. The beginning and the end of each secular resonance (see description in text) is marked by vertical lines, and the resonances are numbered in the middle panels: 1. the inner 3/2 secular resonance, 2. Quasi-Kozai-Lidov resonance, 3. the outer 3/2 secular resonance.}
\label{rev8}
\end{figure}

Figure \ref{rev8} shows a simulation of the evolution of an Earth-Moon system with an initial obliquity of $\theta=65^{\circ}$ through the LPT using our improved numerical integrator. The initial AM of the Earth-Moon system was 2.37 times the current value, so that the ecliptic AM is the same as it is now.  In this example we used constant tidal parameters for Earth and the Moon, with $Q_E/k_{2E}=Q_M/k_{2M}=100$, where $k_2$ and $Q$ are respectively the tidal Love number and the tidal quality factor \citep{md99}, and subscripts E and M refer to Earth and the Moon.  At the end of the simulation, Earth has a low obliquity $\theta=5^{\circ}$, the Moon has an inclination of just over $30^{\circ}$, and the AM is close to that of the present-day system. The small excess in total AM is associated with the lunar inclination which can be damped during the lunar Cassini State transition \citep{che16, cuk16} 

\begin{figure}[ht]
\epsscale{.8}
\plotone{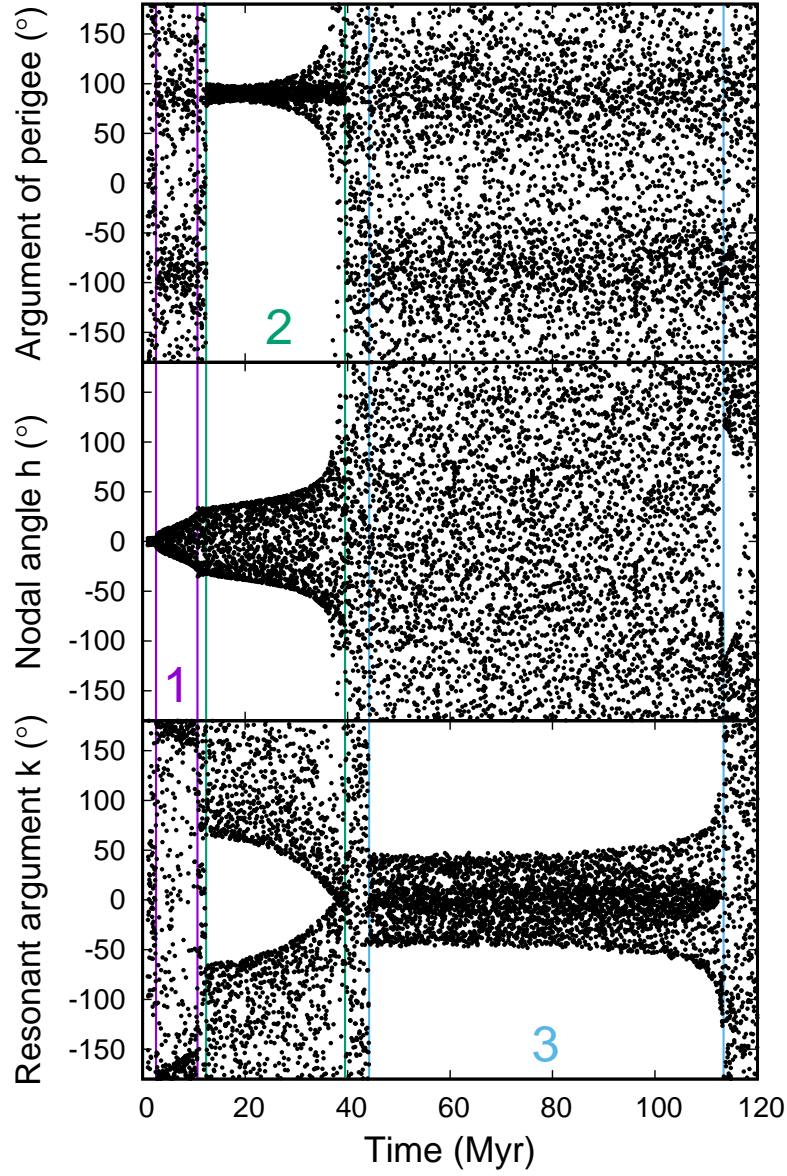}
\caption{The angular variables in the simulation shown in Fig.~\ref{rev8}. The top panel shows the lunar argument of perigee w.r.t. the ecliptic $\omega$. Librations around $\omega=90^{\circ}$ during the interval "2" identify the second of the three secular resonances as a quasi-Kozai-Lidov (QKL) resonance. Middle panel shows the angle between the lunar ascending node (w.r.t. ecliptic) and Earth's vernal equinox $h=\Omega-\Omega_{\rm spin}$. Transition from libration to circulation at about 40 Myr clearly breaks the QKL resonance, but is certainly not the end of the LPT. The bottom panel shows the resonant argument of the outer 3/2 secular resonance $k=3 \Omega -3 \Omega_{\rm spin} + 2 \omega$, which librates during the interval 3, coinciding with the eccentricity excitation and obliquity damping seen in Fig.~\ref{rev8}. Librations of $k$ around $180^{\circ}$ during the QKL resonance are simply a consequence of $k$ being a linear combination of $h$ and $\omega$, which are both librating at this time. Libration of $h$ (in the middle panel) at the very end of the simulation is due to the amplitude of Earth's nutation being larger than that of precession.}
\label{angles}
\end{figure}

There are two distinct phases to the LPT in Fig. \ref{rev8}. Before about 40 Myr into the simulation, the node of the lunar orbit with respect to the ecliptic is librating around the line of the equinoxes, rather than circulating (Fig. \ref{angles}, middle panel). At 40 Myr, when the lunar inclination with respect to the ecliptic reaches zero, the lunar line of nodes stops being tied to the line of equinoxes and starts circulating. This transition was identified as critical by \citet{tia20} and they referred to this event as the termination of the LPT. While an important dynamical event, we must point out that the timing of the transition from libration to circulation of the angle between the lines of lunar nodes and Earth's equinoxes (which we designate $h=\Omega - \Omega_{\rm spin}$ after \citet{tia20}) depends not only on the usual determinants of the Laplace Plane (Earth's mass, obliquity, oblateness etc.) but also on the Moon's free inclination. The circulation in $h$ simply indicates that the tilt of the lunar Laplace Plane to the ecliptic (i.e. lunar forced inclination) is now smaller than the Moon's free inclination. However,  the Laplace Plane Transition is certainly not over, as the Moon's instantaneous Laplace plane at this point is still inclined to the ecliptic by about 30$^{\circ}$ (red line, Fig. \ref{rev8}).       

The transition between libration and circulation in $h$ leads to a temporary drop in lunar eccentricity (Fig. \ref{rev8}, middle left panel). Excitation of eccentricity is generally due to secular resonances present during the LPT, and this event represents the end of the second of three major secular resonances that dictate the dynamics of the orbital evolution shown in Fig. \ref{rev8}. We discuss the other secular resonances below. This middle resonance lasts from about 13 Myr to 40 Myr, and the top panel in Fig. \ref{angles} shows that during this period (labeled "2" in Figs. \ref{rev8} and \ref{angles}) the lunar argument of perigee with respect to the ecliptic, $\omega$, is librating, indicating a form of Kozai-Lidov resonance \citep{lid62, koz62}. Unlike in the usual Kozai-Lidov problem, where the outer perturber dominates, here the precession is mostly driven by the Earth's oblateness. While $h$ librates, the argument of pericenter referenced to the ecliptic is $\omega=\varpi-\Omega_{\rm spin}$, where $\varpi$ and ${\Omega_{\rm spin}}$ are the longitudes of the lunar perigee and Earth's vernal equinox, respectively. Since $\dot{\Omega}_{\rm spin} << \dot{\Omega}_{\rm eq}$ (where $\Omega_{\rm eq}$ is the lunar node with respect to Earth's equator), this ``Quasi-Kozai-Lidov'' (QKL) resonance requires that $\dot{\omega}_{\rm eq} \approx -\dot{\Omega}_{\rm eq}$, a kind of secular resonance previously studied in the context of artificial satellite dynamics \citep{ros15}. The Sun, through the Kozai-Lidov mechanism keeps exciting the lunar eccentricity, effectively removing AM from the system, while preserving the ecliptic AM. Given the criteria for this resonance, it will necessarily break once the nodal parameter $h$ starts circulating (middle panel in Fig. \ref{angles}).

The QKL resonance found in our simulations differs from other variations of the Kozai-Lidov resonance that are observed when there is also an inner perturber present in the system (the original Kozai-Lidov resonance relies exclusively on the outer perturber). When there is little or no obliquity of the central body (or the inner perturber is coplanar with the outer), the definition of the argument of pericenter are the same in equatorial and ecliptic reference frames. In that case, a single threshold for instability (i.e., excitation of eccentricity for initially circular orbits) connects Molniya orbits at 63.4$^{\circ}$ (close to the planet) and the Kozai-Lidov threshold at 39.2$^{\circ}$ far from the planet \citep{tre14, pet20}. Since our inner perturber (Earth) has a large obliquity, our secular dynamics is more complex. QKL resonance in the present paper relies on the longitude of the node being almost stationary, as the orbital plane is dominated by the tilt of the planet's equatorial plane. In this configuration, the argument of pericenter (with respect to the Sun) precesses at rates similar to the longitude of pericenter, and the stationary $\omega$ with respect to the Sun can be reached at lower free inclinations than in the low obliquity case. Once $h$ begins to circulate due to decreasing obliquity, this form of instability disappears without free inclination crossing any critical values. Similarly, while our QKL resonance is in the regime where precession is dominated by Earth's oblateness, the resonant interaction is with the Sun (as our Earth is azimuthally symmetric). This sets the QKL apart from the outer Kozai-Lidov resonance \citep{nao17} in which the resonance arises with interactions with an eccentric inner perturber. 

After a brief period ($\sim5$~Myr in the example in Fig. \ref{rev8}) of low inclination, the system enters another secular resonance. This final resonance is absent in the results of \citet{tia20} and makes our simulated evolution diverge fundamentally from theirs.  Lasting from about 44 Myr to 114 Myr (labeled "3" in Figs. \ref{rev8} and \ref{angles}), this secular resonance excites lunar eccentricity and therefore arrests further outward tidal evolution. In this dynamical state, the lunar inclination remains approximately constant ($i \approx 33^{\circ}$), while Earth's obliquity decreases, while conserving ecliptic AM (Fig. \ref{rev8}). We find that this resonance is caused by the periods of variation of lunar inclination (due to perturbations from Earth's oblateness) and lunar eccentricity (due to the Kozai-Lidov mechanism) being exact multiples of each other. For each complete cycle in lunar inclination (driven by the circulation of the nodal angle $h$) there are three cycles in eccentricity (dependent on the angle $2 \omega$). Therefore, the resonant argument is $k=3h+2\omega$ or $k=3 \Omega -3 \Omega_{\rm spin} + 2 \omega$. This secular resonance requires that $\dot{\omega} \approx - (3/2) \dot{\Omega}$, which is consistent with the lunar inclination in our simulations (see Discussion). The bottom panel in Fig. \ref{angles} plots the resonant argument $k$ during the simulation shown in Fig. \ref{rev8}, with libration of $k$ confirming the presence of the resonance. This resonance was found by \citet{cuk16}, but its effects were unfortunately entangled with the erroneous obliquity drift in their simulations. Our corrected simulations here confirm that this secular resonance takes non-ecliptic AM from Earth's rotation and, through the Kozai-Lidov mechanism, passes it to the heliocentric orbit of Earth. Fig. \ref{rev8} shows that this resonance can last until Earth's obliquity is almost completely removed, at which point the resonance breaks.  

\begin{figure}
\plotone{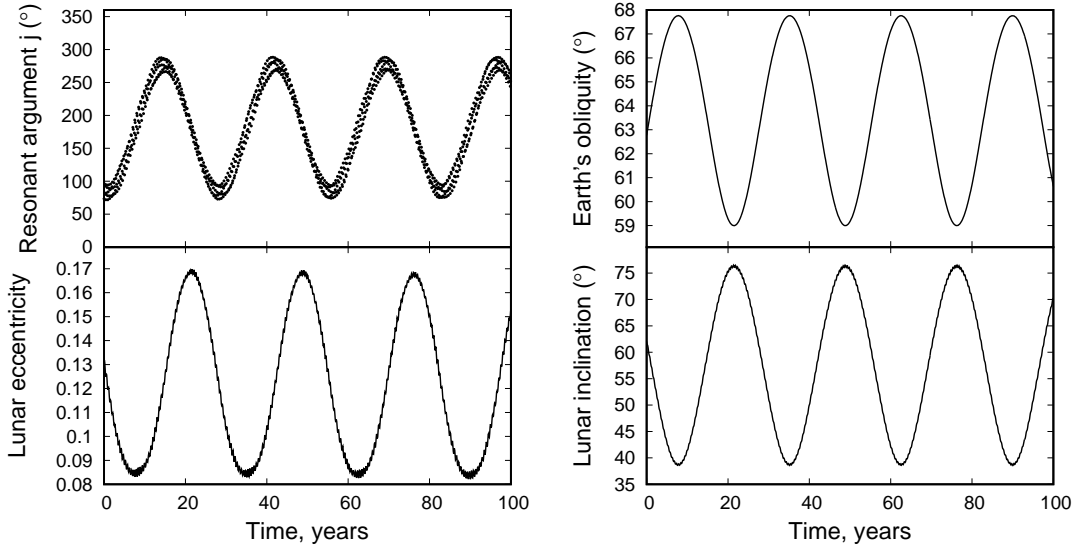}
\caption{A short simulation illustrating the state of the system 10 Myr into the simulation shown in Fig. \ref{rev8}, illustrating the inner 3/2 secular resonance (Table 1). The top left panel plots the resonant argument $j=3 \Omega_{eq} + 2 \omega_{eq} - 2 \Omega_{spin}$, where $\Omega_{eq}$ and $\omega_{eq}$ are the lunar longitude of the ascending node and argument of perigee in the equatorial reference frame, while $\Omega_{spin}$ is the longitude of vernal equinox. Left-hand panels show Earth's obliquity and lunar inclination relative to the ecliptic.}
\label{first1}
\end{figure}

\begin{table}
\centering
\caption{Secular resonances identified in the simulation plotted in Fig. \ref{rev8}} 
\begin{tabular}{llccc}
\hline
\# in Fig. \ref{rev8} & Time in Fig. \ref{rev8} & Resonant argument & \# of $e$ cycles per $i$ cycle & Designation \\
\hline
1 & 3-11 Myr &  $j=3 \Omega_{eq} + 2 \omega_{eq} - 2 \Omega_{spin}$ & 1 & Inner 3/2 secular resonance  \\
2 & 13-40 Myr & $2 \omega$ & 2 & Quasi-Kozai-Lidov (QKL) resonance \\
3 & 44-114 Myr &  $k=3 \Omega -3 \Omega_{spin} + 2 \omega$ & 3 & Outer 3/2 secular resonance \\
\hline
\end{tabular}
\end{table}

The earliest, short-lived secular resonance seen within the first 10 Myr in Fig. \ref{rev8} (labeled "1") is related to the late secular resonance we just described, likewise requiring the precession rate of the argument of perigee to be about -150\% of that of the node. However, these orbital elements need to be measured relative to Earth's equator, rather than the ecliptic,  as the perturbations from Earth's oblatenesss still dominates the lunar orbit at this early time. The relevant resonant argument is $j=3 \Omega_{\rm eq} + 2 \omega_{\rm eq} - 2 \Omega_{\rm spin}$. Note that $j$ does not obey D'Alembert's rules \citep{md99} because it is written using variables from two different coordinate systems\footnote{In contrast, $k$ does satisfy D'Alembert's rules, as the argument of pericenter $\omega=\varpi-\Omega$ is in itself a difference of two longitudes.}. Figure \ref{first1} shows a short snapshot of the first secular resonance 10 Myr into the simulation, illustrating the libration of $j$ and the coupling between different dynamical quantities. To avoid confusion, from here on we will refer to the first and third secular resonances encountered during our example lunar orbital evolution as the inner and outer 3/2 secular resonances, respectively, while we will continue to refer to the second resonance as the quasi-Kozai-Lidov (QKL) resonance (see Table 1).

\section{High Eccentricity Episodes}\label{sec:highe}

\begin{figure}
\epsscale{.6}
\plotone{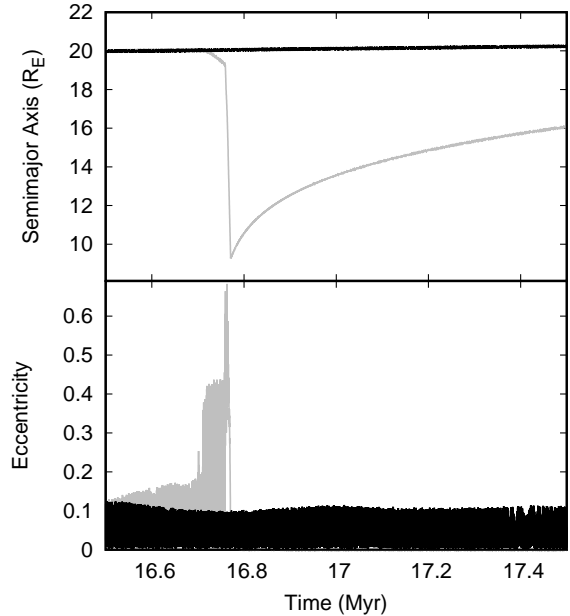}
\caption{Part of an integration that experienced rapid collapse of semimajor axis. Gray line plots the initial simulation with initial obliquity of $\theta=62^{\circ}$, the correct ecliptic AM, and $Q_M/k_{2M}=Q_E/k_{E2}=100$. At the point in orbital evolution when the quasi-Lidov-Kozai resonance is encountered by other simulations, this integration appears to be captured into a very high-$e$ secular resonance, leading to rapid collapse of semimajor axis. Changing the lunar tidal dissipation to $Q_M/k_{2M}=50$ (black line) avoids this instability. We argue that this instability is probably not relevant for the real Earth-Moon system, as it requires the Moon to maintain the same tidal parameters despite having very high eccentricity and so experiencing intense tidal heating.}
\label{collapse}
\end{figure}

The sequence of secular resonances encountered in the simulation shown in Fig. \ref{rev8}, ending with a low-obliquity Earth, is typical of most of our integrations that have explored starting with a compact Earth-Moon system that has the present-day ecliptic AM and obliquities of $62^{\circ}$ or $65^{\circ}$, with various tidal properties of Earth and the Moon. For some integrations restricted to initial obqliuities of $61^{\circ}-62^{\circ}$, relatively early in the LPT the lunar orbit is captured in a secular resonance that forces lunar eccentricity to become exceptionally high ($e>0.5$), as seen in Fig. \ref{collapse}. In such scenarios, extremely strong satellite tides quickly collapse the lunar semimajor axis, leading to simulations crashing as our fixed timestep cannot handle much smaller $a$. When repeated with a smaller timestep, simulations do not suffer crashes, but the Moon effectively re-starts its tidal evolution at a smaller $a$ with completely damped eccentricity and inclination. As some of the total AM has already been lost and ecliptic AM is conserved, Earth's obliquity is now lower than it was initially, and the system can evolve without encountering the Laplace Plane instability. The end result would be a high AM, high obliquity system with a non-inclined Moon, inconsistent with the present-day Earth-Moon system.

We do not expect such an evolutionary track is a realistic possibility for the Earth-Moon system. An increase in the tidal dissipation within the Moon avoids the secular resonance found to produce very high eccentricities, resulting in an orbital evolution like the one shown in Fig. \ref{rev9} and leading to a low-obliquity Earth. High eccentricity leads to intense tidal heating of the Moon and, as shown by \citet{tia17}, tidal heating affects the Moon's tidal properties, typically leading to the increase in the parameter $A$, which measures the relative strength of tides within the Moon and Earth \citep{can99}:
\begin{equation}
A={k_{2M}Q_E M_E^2 R_M^5 \over k_{2E} Q_M M_M^2 R_E^5}
\end{equation}
where $M$ and $R$ are mass and radius, and subscripts E and M refer to Earth and the Moon. This dependence of tidal properties on tidal dissipation would act to moderate or eliminate high-$e$ episodes of lunar tidal evolution \citep{tia17}. Therefore, we believe that cases of lunar semimajor axis collapse due to high eccentricities are an artifact of our use of constant tidal properties for the Moon, rather than a plausible orbital evolution pathway. However, we think that the use of constant tidal properties is generally justified in other cases, such as evolutions shown in Figs. \ref{rev8} and \ref{rev9}, as the Moon does not reach very high eccentricities during those simulations.

The high-eccentricity evolution shown here is restricted to cases with Earth's initial obliquities in the $61^{\circ}-62^{\circ}$ range. For $\theta_{initial}=61^{\circ}$ we find instability using both $A=10$ and $A=20$, while for $\theta_{initial}=62^{\circ}$ high-$e$ crash happens only when $A=10$. Apart from the above discussion of how variable tidal properties may affect this instability, we note that \citet{tia20} integrated systems with $\theta_{initial}=61^{\circ}$ without suffering a high-$e$ crash.  Clearly this phenomenon is at least somewhat dependent on the tidal model used, and therefore more work is needed to evaluate its intrinsic relevance to the evolution of a high-obliquity Earth-Moon system.

\section{Obliquity of Earth after the Laplace Plane Transition}\label{sec:obl}

The evolution through the three secular resonances plotted in Fig. \ref{rev8} produces a system with an AM similar to that of the present-day Earth-Moon system but with a very low obliquity of Earth. Further tidal evolution would have made Earth's obliquity increase as the Moon absorbed more and more ecliptic AM from Earth \citep{tou94}. Based on the constraints of the current lunar orbit and terrestrial rotation rate, the post-LPT obliquity of Earth would need to have been in the range $10-15^{\circ}$ \citep{tou94, rub16}, somewhat higher than that found at the end of the simulation in Fig. \ref{rev8}. Simple extrapolation from work of \citet{tou94} would imply that this simulations would lead to present-day Earth obliquity of $\le 10^{\circ}$. We find that the dynamics of the end of the LPT are quite complex and dependent on the tidal properties of Earth and the Moon and that different choices of tidal properties can lead to systems with higher final obliquities. 

\begin{figure}
\plotone{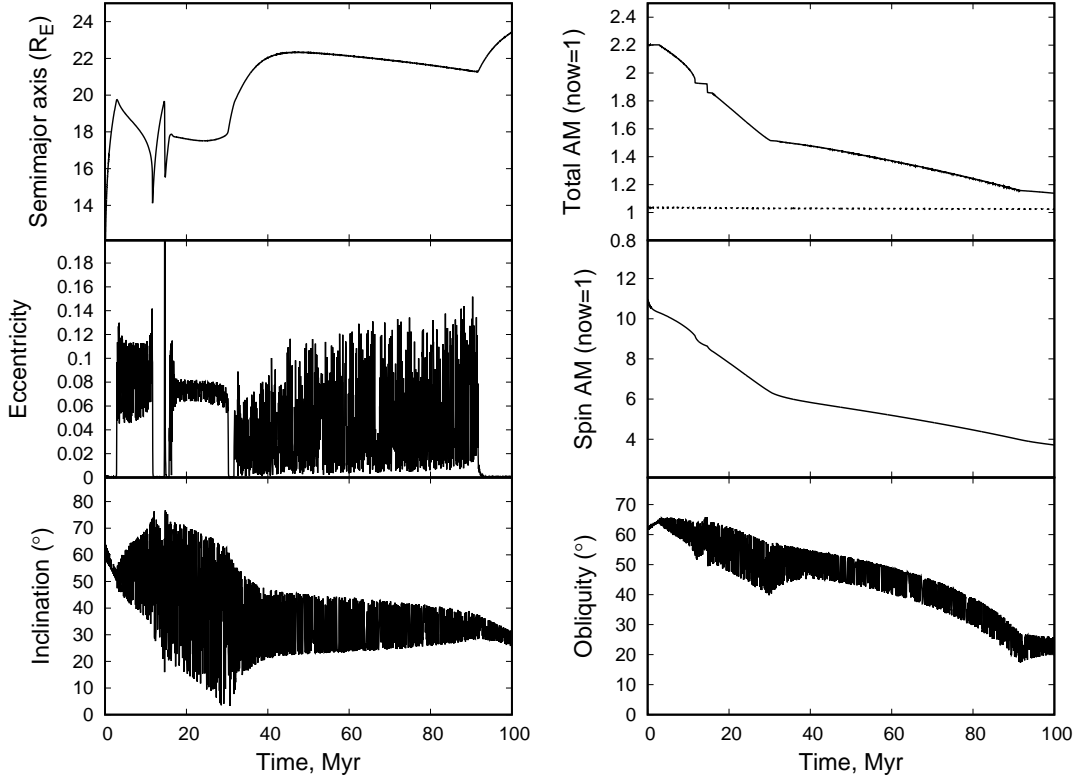}
\caption{Evolution of the Earth-Moon system starting with a 62$^{\circ}$ obliquity and 2.13 times its current total AM. We used constant tidal parameters $Q_E/k_{2E}=100$ and $Q_M/k_{2M}=50$ for Earth and the Moon, respectively. The panels plot the same quantities as in Fig. \ref{rev8}. The main difference from the simulation in Fig. \ref{rev8} is that the Moon's orbit is never librating within the outer 3/2 secular resonance, but is circulating outside the resonance (Fig. \ref{end_rev9}), resulting in a higher obliquity of Earth ($\theta\approx 20^{\circ}$) at the end of the simulation.}
\label{rev9}
\end{figure}

\begin{figure}
\plotone{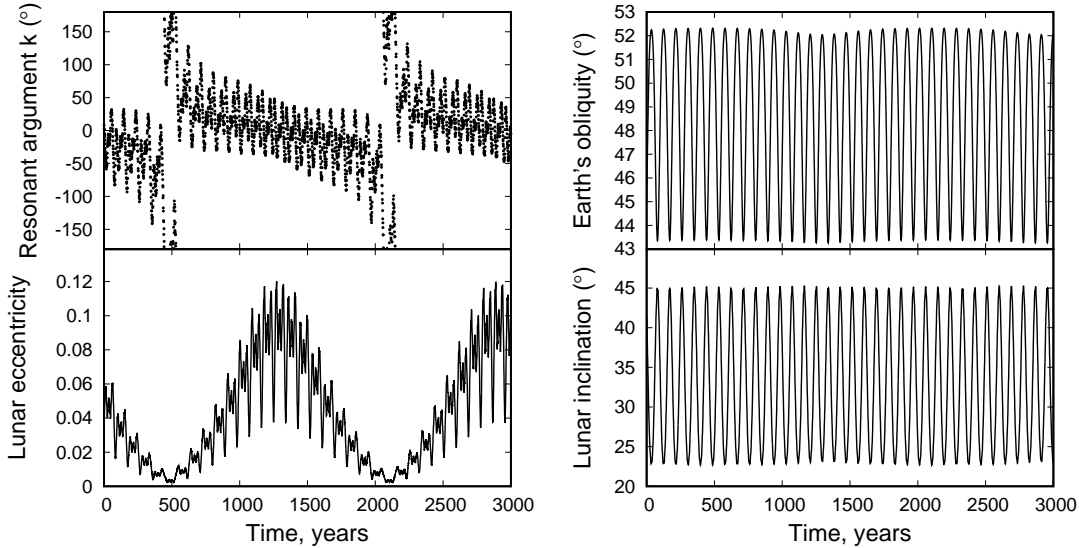}
\caption{A short integration illustrating the dynamical state of the system 50 Myr into the simulation shown in Fig. \ref{rev9}. The top left panel plots the resonant argument $k=3 \Omega -3 \Omega_{\rm spin} + 2 \omega$, where $\Omega$ and $\omega$ are lunar longitude of the ascending node and argument of perigee, and $\Omega_{\rm spin}$ is the longitude of the vernal equinox (in an ecliptic reference frame). Other three panels plot Earth's obliquity to ecliptic (top right), and the Moon's eccentricity (bottom left) and inclination relative to the ecliptic (bottom right). While the system spends most of the time at $k \approx 0^{\circ}$, the argument $k$ is circulating, rather than librating as in Fig. \ref{angles}.}
\label{end_rev9}
\end{figure}

Figure \ref{rev9} shows a tidal evolution similar to that in Fig. \ref{rev8}, except that the Earth begins with somewhat lower obliquity and AM (although with the same ecliptic AM), and that the lunar tidal response is twice as strong ($A=20$ vs. $A=10$). The initial conditions do not seem to affect the final outcome, as simulations with the current ecliptic AM tend to pass through a ``keyhole'' in phase space at the end of the QKL resonance (labeled ``2'' in Fig. \ref{rev8} and \ref{angles}), with a total AM about 50\% greater than the present-day and obliquity $\theta \approx 50^{\circ}$. This ``keyhole'' is equivalent to the end of LPT proposed by \citet{tia20}, except that in our simulations it is followed by the outer 3/2 secular resonance. Fig. \ref{rev9} shows that this resonance is significantly affected by the lunar tidal response, with lunar eccentricity periodically going through zero, indicating a lack of libration. Figure \ref{end_rev9} shows a snapshot of this near-resonance at 50 Myr into the simulation shown in Fig. \ref{rev9}. Fig. \ref{end_rev9} indicates that the same resonant angle $k$ as in Fig. \ref{angles} is responsible for the excitation of lunar eccentricity, despite not being in libration. Given that the system is not as deep in the outer 3/2 secular resonance in Fig. \ref{rev9} as in Fig. \ref{rev8}, the Moon exits the resonance when Earth's obliquity is about $\theta \approx 20^{\circ}$. This value is probably too high to match the present Earth-Moon system and would result in $30^{\circ}$ obliquity for Earth. However, a full numerical study of the tidal evolution out to $60 R_E$ \citep[updating ][]{tou94} is probably necessary in order to quantify the correspondence between post-LPT obliquities and the current day values for Earth-Moon system histories with large-scale lunar inclination damping. 

The lowest initial obliquity for which the Laplace Plane instability occurs is about $\theta=61^{\circ}$, while the upper limit of obliquities is determined by the plausible upper limit on the total AM of the post-giant-impact Earth. Obliquities larger than $\theta \approx 72^{\circ}$ would require initial AM that are typically not found in the final states of numerical giant impact simulations \citep{Lock2017,loc18,cuk12,can12,loc20}. 

Our integrations indicate that the final obliquity of Earth depends on the tidal parameters more than the initial obliquity and AM. In particular, early indications are that the tidal response of Earth and the Moon during and after capture of the Moon into the outer 3/2 secular resonance determines the final obliquity of Earth. The dynamics at resonance capture determine whether further evolution will involve libration (Fig. \ref{angles}) or borderline circulation (Fig. \ref{end_rev9}) of the resonant argument $k$. We will explore the sensitivity of the evolution to tidal parameters further in future work. The fact that the reconstructed obliquity of Earth \citep{tou94} is within the range of our early results (Figs. \ref{rev8}, \ref{rev9}) gives us confidence that a high-obliquity pathway to today's exact configuration will be found. 

\section{Discussion}\label{sec:dis}

In the previous sections, we have demonstrated that an initially high-obliquity, high-AM Earth-Moon system ($\theta > 61^{\circ}$) naturally evolves into a state with much lower AM, a low obliquity of Earth and a lunar inclination of $i \approx 30^{\circ}$. We confirm the finding of \citet{tia20} that the ecliptic AM is conserved during the LPT and agree with them that the original simulations of \citet{cuk16} suffered from significant numerical errors. However, our findings of low final obliquity and AM, consistent with the present Earth-Moon system, directly contradict the conclusions of \citet{tia20}  that the Laplace Plane instability cannot lead to the system's present configuration. 

Comparing our simulations with those of \citet{tia20}, it is evident that the crucial difference in our simulations is the presence or absence of the dynamical feature we label the outer 3/2 secular resonance, in which the resonant argument $k$ either librates (Fig. \ref{angles}) or is at the boundary between circulation and libration (Fig. \ref{end_rev9}). In order to understand when this resonance is encountered, we explored where it is located in orbital element space. While the dynamics of a satellite in the LPT can be very complex \citep{tre09, tam13}, Figures \ref{rev8} and \ref{rev9} show that the outer 3/2 secular resonance is encountered at a point where the Sun is the principal perturber on the lunar orbit. This is the well-known Kozai-Lidov regime \citep{lid62, koz62} and we can use the following expressions for the orbital precession rates \citep{inn97, car02}:
\begin{equation}
{d\omega \over d\tau} = {3 \over 4} [2 (1-e^2) + 5 \sin^2 \omega (e^2 - \sin^2 i)]/\sqrt{1-e^2}
\end{equation}
\begin{equation}
{d\Omega \over d\tau} =  - {\cos{i} \over 4} [3 + 12 e^2 -15 e^2 \cos^2{\omega}]/\sqrt{1-e^2}  
\end{equation}
where $\tau$ is dimensionless time, and all orbital elements are in an ecliptic coordinate system. Since $e \simeq 0.1$ during much of the duration of the outer 3/2 secular resonance, while $\sin{i} \simeq 0.5$, we will ignore terms involving $e^2$ but keep those with $\sin^2{i}$. We will also assume that $\omega$ is circulating faster than the resonant argument $k$ (Figs. \ref{angles} and \ref{end_rev9}), so that $<\sin^2{\omega}> =0.5$, which is not true during Kozai-Lidov resonance but is acceptable here, as $\omega$ has to circulate when $h$ is circulating and $k$ is librating. Therefore, if we assume that Earth's axial precession is much slower than the Moon's orbital precession, the condition for the libration of $k$ becomes $\dot{\omega}= - (3/2) \dot{\Omega}$. Combining the above equations, we find that $5 \cos^2{i} - 3\cos{i} -1=0$. The solution to this equation is $\cos{i}=(3+\sqrt{29})/10$, or $i \approx 33^{\circ}$. This result matches the average inclination of the Moon while in the outer 3/2 secular resonance in Figs. \ref{rev8} and \ref{rev9} rather well. 

The lunar inclination at the end of the simulations shown in \citet{tia20} Fig. 2 is only slightly larger than that expected during the outer 3/2 secular resonance, $i \approx 34^{\circ}$. While this difference in inclination is apparently sufficient to avoid the secular resonance, the Moon has a similarly high inclination in our integrations at the end of the QKL resonance and therefore this cannot be the cause of the different results of our studies. All our simulations that feature the QKL resonance have the Moon subsequently experience either resonant or near-resonant evolution along the 3/2 outer secular resonance, evolving the Earth to low obliquities. However, in the simulations of \citet{tia20}, the Moon avoids the 3/2 outer secular resonance because its inclination remains above $33^{\circ}$ despite the expectation of inclination damping \citep{che16}. The limited inclination damping is the result of \citet{tia20} using a constant figure of the Moon, based on the hydrostatic shape at the distance of 6 $R_E$. This assumption makes the Moon unrealistically non-spherical at the end of QKL resonance, and makes the model of \citet{tia20} underestimate by orders of magnitude the lunar obliquity tides and the extent of the resulting inclination damping. 

The lunar inclination in our Figs. \ref{rev8} and \ref{rev9} is constant only while the Moon is in the outer 3/2 secular resonance. Once the resonance is broken (at the very end of both simulations), lunar inclination decreases noticeably over time. This decrease is due to obliquity tides and is rather dramatic as we assumed strong lunar dissipation in these simulations ($A=10$ and $A=20$ similar to that used by \citet{tia20}). Inclination damping is related to the tidal evolution rate as \citep{chy89}:
\begin{equation}
{1 \over \sin{i}} {d\sin{i} \over dt} = - {A \over 2} {\dot{a} \over a} \left( {\sin{\theta_M} \over \sin{i}} \right)^2    
\end{equation}
where $\theta_M$ is the lunar obliquity, and we assumed that $\dot{a}$ is dominated by tides raised on Earth by the Moon. In our simulations, the Moon is in Cassini state 1 around $a=23 R_E$, and $(\theta_M/i) \approx 0.25$. Using analytical estimates for the lunar figure \citep{gar06}, axial precession \citep{war75} and both solar-induced \citep{inn97} and oblateness-induced \citep{dan92} nodal precession, we find axial and nodal precession periods of about 20 yr and 80 yr, respectively, consistent with this $\theta_M/i$ ratio. Therefore, we would expect $\sin{i}^{-1} (di/dt) \approx - 0.3 (\dot{a}/a)$, so when the Moon evolves from 18 to 24$R_E$ at the end of simulations in Fig. 2 of \citet{tia20}, we would expect lunar inclination to decrease noticeably, by several degrees, which is not seen in their simulations. Note that this calculation does not depend on assumptions about lunar shape, as the Moon's current oblateness and the oblateness expected from hydrostatic equilibrium are equal at about $23 R_E$.

\begin{figure}
\plotone{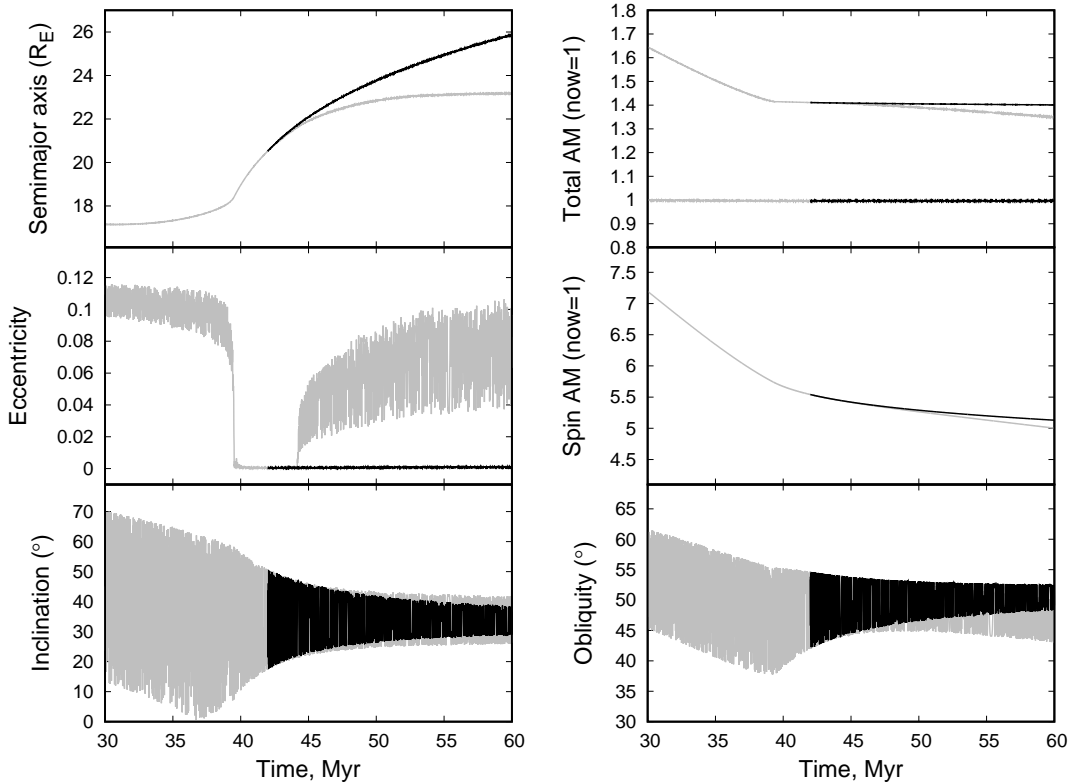}
\caption{A simulation of orbital evolution by a constant-figure Moon \citep[as used by][]{tia20}, plotted in black, branching at 42~Myr from the simulation shown in Fig. \ref{rev8} (gray). The panels plot the same quantities as in Fig. \ref{rev8}. As in the simulations of \citet{tia20}, the Moon does not encounter the outer 3/2 secular resonance as it evolves outward on a circular orbit.}
\label{shtest42}
\end{figure}

To illustrate the importance of taking into account the correct lunar figure, we ran a control simulation using the same fixed lunar shape as \citet{tia20}. For initial conditions we used the state of the system 42~Myr into the simulation shown in Fig. \ref{rev8}, which is between residence in the QKL and the outer 3/2 resonances. Results are plotted in Fig. \ref{shtest42}; the Moon maintains a constant orbital inclination and does not encounter outer 3/2 resonance as it moves out beyond 25~$R_E$.

We tentatively conclude that the main difference between our model and that of \citet{tia20} is their apparent lack of significant inclination damping due to obliquity tides upon leaving the QKL resonance, which makes the encounter with the outer 3/2 secular resonance much less likely. In our integrations shown in Figs \ref{rev8} and \ref{rev9}, the lunar inclination is above $i=33^{\circ}$ when the QKL resonance is broken, as the QKL resonance requires slower $\dot{\omega}$ (and therefore higher $i$) than the 3/2 outer secular resonance. Through the action of obliquity tides, the lunar inclination steadily decreases and enables capture into (or just outside) the outer 3/2 secular resonance. We note that this resonance is present in some of \citet{tia20} simulations included in their Supplementary material. Since the outer 3/2 secular resonance is driven primarily by solar perturbations, it is not finely sensitive to Earth's spin and is therefore also present in cases with a lower Earth-Moon system AM. In \citet{tia20} Fig. S1, the eccentricity of the Moon is weakly excited after $h$ starts circulating, most likely because of proximity to the outer 3/2 secular resonance. While the effect on the total AM is small, there is a noticeable decrease in Earth's obliquity while the resonance is active. Additionally, some of the final obliquities of Earth in Fig. S3 of \citet{tia20} are notably smaller than the corresponding lunar inclinations, indicating that some obliquity-reducing mechanism must have been acting after $h$ started circulating. 

The evolution of the Earth-Moon system's angular momentum, Earth's obliquity, lunar inclination, lunar tidal response and lunar shape are all closely connected when the Moon is at $20-25 R_E$. A deformable, and therefore highly dissipative, Moon is required for AM loss and a reduction in Earth's obliquity through the outer 3/2 secular resonance. However, a continuing strong tidal response of the Moon as it approaches the Cassini state transition \citep[at 30-34$R_E$;][]{war75, cuk16, cuk19} would completely remove all of the lunar inclination, which is in conflict with the present day lunar orbital tilt \citep{che16}\footnote{Under some assumptions about lunar tidal evolution, encounters between the Moon and late-surviving planetesimals could generate lunar inclination after the LPT \citep{pah15}. \citet{cuk16} discuss some limitation of this mechanism.}. \citet{cuk16} showed that a Moon with close to present-day tidal properties can damp the lunar inclination from values around  $i=30^{\circ}$ at $25 R_E$ to the present day value, $i=5^{\circ}$. Therefore, the Moon must have transitioned from a very strong tidal response (expected from a fluid body) to a state close to its current solid-body response around $25 R_E$, or soon after the end of the outer 3/2 secular resonance, according to our results. We also note that the Earth-Moon distance at which the Moon's current shape was frozen in is likely in the $a=23-30 R_E$ range \citep{gar06}, but the question of lunar shape is notoriously complex, and any connection to LPT will need to be studied separately. In any case, our model makes definite predictions about the Moon's early orbital, geophysical and thermal evolution, and we hope that further work on lunar geological history and paleomagnetism will be able to further constrain early lunar tidal evolution.

Finally, the requirement of initial obliquity in the $61-71^{\circ}$ range is actually more likely by a factor of $\approx 5$ (assuming random distribution on a sphere) than the initial obliquity of $<10^{\circ}$. The canonical model \citep{can01} requires $\theta_0=10^{\circ}$, but even lower obliquities are needed to match AM loss through the evection resonance \citep{cuk12}. The constraint of conservation of ecliptic AM requires that systems with obliquities of $61-71^{\circ}$ have total initial AM of 2.1-3.1 times the present-day AM of the Earth-Moon system. It has been shown that many such high-AM impacts can produce post-impact structures with significant amounts of mass and AM beyond the Roche limit and with silicate vapor pressures at the Roche-limit of tens of bars, consistent with the moderately-volatile depleted composition of the Moon \citep[see Figures~14 and 17 in][]{loc18}.

\section{Conclusions}\label{sec:con}

We confirm the finding of \citet{tia20} that the ecliptic angular momentum of the Earth-Moon system is approximately conserved during the Laplace-plane transition. However, contrary to their claims, we find that Earth-Moon system with a high initial obliquity ($\theta > 61^{\circ}$) and AM (combined to produce the correct ecliptic AM) naturally evolves into a state with the current total AM and a low obliquity of Earth. The crucial dynamical mechanism that allows for the evolution to low obliquities is a secular resonance (the ``outer 3/2 secular resonance'') that is invariably encountered by the (then highly inclined) lunar orbit during the later stages of the Laplace Plane transition. The capture into this secular resonance requires some inclination damping, which is expected for realistic obliquity tides, but is absent in the results of \citet{tia20} due to their treatment of the lunar figure.

We conclude that the hypothesis of an initially highly tilted Earth with a high AM is viable and offers much promise in explaining the implied common source for terrestrial and lunar materials \citep{cuk12,can12,loc18}, the moderately volatile depleted composition of the Moon \citep{loc18}, and subsequent AM loss. It also naturally explains the current lunar inclination while allowing for large scale core-mantle friction during the lunar Cassini State transition, potentially powering the ancient lunar dynamo \citep{cuk19}. 
\begin{acknowledgments}
M\'C is supported by NASA Emerging Worlds Program award 80NSSC19K0512. SJL gratefully acknowledges funding from NSF through award EAR-1947614. Comments by Daniel Tamayo and two anonymous reviewers greatly improved the  manuscript. The authors would like to thank Jihad Touma and the faculty and staff of the American University in Beirut for hosting a workshop on lunar formation in October 2019, which greatly advanced our understanding of early lunar evolution.
\end{acknowledgments}

\bibliography{cuk_refs}{}
\bibliographystyle{aasjournal}



\end{document}